\documentclass[aps,prl,superscriptaddress,twocolumn,amsmath,amssymb,floatfix]{revtex4-1}
\usepackage{tabularx}
\usepackage{bm}
\usepackage{euscript}
\usepackage{graphicx}
\usepackage{color}
\usepackage{amsfonts}
\usepackage{exscale}
\usepackage{amsbsy}
\usepackage{textcomp}
\usepackage{dcolumn}
\usepackage{amsmath}
 \usepackage{xfrac}
 \usepackage{url}

\def\be{\begin{equation}}
\def\ee{\end{equation}}
\def\ba{\begin{eqnarray}}
\def\ea{\end{eqnarray}}

\def\SRO{Sr$_2$RuO$_4$}
\def\URS{URu$_2$Si$_2$}

\begin{document}

\title{Evidence for broken time-reversal symmetry in the superconducting phase of \URS}

\author{E.~R.~Schemm}
\affiliation{Department of Physics, Stanford University, Stanford, CA 94305, USA}
\affiliation{Stanford Institute for Materials and Energy Sciences, SLAC National Accelerator Laboratory, 2575 Sand Hill Road, Menlo Park, CA 94025, USA}
\author{R.~E.~Baumbach}
\altaffiliation{Present address: National High Magnetic Field Laboratory, Tallahassee, FL 32310, USA}
\affiliation{Los Alamos National Laboratory, Los Alamos, NM 87545, USA}
\author{P.~H.~Tobash}
\affiliation{Los Alamos National Laboratory, Los Alamos, NM 87545, USA}
\author{F.~Ronning}
\affiliation{Los Alamos National Laboratory, Los Alamos, NM 87545, USA}
\author{E.~D.~Bauer}
\affiliation{Los Alamos National Laboratory, Los Alamos, NM 87545, USA}
\author{A.~Kapitulnik}
\email[Correspondence to: ]{aharonk@stanford.edu}
\affiliation{Department of Physics, Stanford University, Stanford, CA 94305, USA}
\affiliation{Stanford Institute for Materials and Energy Sciences, SLAC National Accelerator Laboratory, 2575 Sand Hill Road, Menlo Park, CA 94025, USA}
\affiliation{Department of Applied Physics, Stanford University, Stanford, CA 94305, USA}

\begin{abstract}
Recent experimental and theoretical interest in the superconducting phase of the heavy fermion material \URS~has led to a number of proposals in which the superconducting order parameter breaks time-reversal symmetry (TRS). In this study we measured polar Kerr effect (PKE) as a function of temperature for several high-quality single crystals of \URS. We find an onset of PKE below the superconducting transition that is consistent with a TRS-breaking order parameter. This effect appears to be independent of an additional, possibly extrinsic, PKE generated above the hidden order transition at $T_{HO}=17.5$ K, and contains structure below $T_c$ suggestive of additional physics within the superconducting state.
\end{abstract}

\date{\today}

\maketitle

The heavy fermion semimetal \URS~\cite{Palstra1985,Maple1986,Schlabitz1986} is perhaps the most enigmatic of the $f$-electron compounds owing to its interplay of a so-called hidden order state with antiferromagnetism and superconductivity. While the hidden order phase ($T_{HO}=17.5$ K) has long been the primary  focus of experimental and theoretical studies of \URS~\cite{Mydosh2011}, superconductivity which emerges from the hidden order at lower temperatures ($T_c \approx 1.5$ K), was recognized early on to be intimately linked to the HO phase \cite{Mason1990,Isaacs1990}, and unconventional in nature \cite{Brison1994,Kohori1996,Matsuda1996}. Both orders are quenched by  under uniaxial strain, giving up to a large moment antiferromagnetism phase  \cite{Amitsuka2007,Hassinger2008}.

Renewed experimental and theoretical interest in the superconducting state has elucidated a number of its properties in recent years. Angle-resolved specific heat \cite{Yano2008} and thermal conductivity \cite{Kasahara2007,Kasahara2009a,Kasahara2009b} measurements suggest that \URS~is a chiral $d$-wave, even parity, multi-gap superconductor whose gap function belongs to the $E_g$ representation of the crystal point group $D_{4h}$; the specific form of the gap based on analysis of the nodal structure is argued to be
\be
\Delta(\textbf{k})=\Delta_0 {\rm sin}\frac{k_zc}{2}\left( {\rm sin}\frac{k_x+k_y}{2}a \pm i  {\rm sin}\frac{k_x-k_y}{2}a\right)
\label{gap}
\ee
where the tetrahedral lattice constants are $a$ and $c$ and the coordinate system takes $\hat{z}$ to be parallel to $\hat{c}$. Such an order parameter exhibits point nodes along $\hat{z}$ and line nodes at $k_z = 0$ and $k_z = \pm{2\pi}/{c}$. In addition, the presence of a relative phase between the two order parameter components leads to an overall pair angular momentum and, hence, broken time-reversal symmetry (TRS) with moment along the $\hat{c}$. Very recently, more exotic models for the superconducting state in \URS~\cite{Goswami2013,Hsu2013} have been proposed which also require TRS to be broken in a similar manner. However, such a TRS-breaking  (TRSB) moment, while inferable from complementary measurements, has yet to be experimentally confirmed.

A direct consequence of TRSB of the form given in Eq.~\ref{gap} is the emergence of a finite polar Kerr effect (PKE) below $T_c$ for light incident on the $ab$ plane and traveling along the $c$ direction. The Kerr angle $\theta_K$ is related to the imaginary component of the 3D Hall conductivity tensor $\sigma_{xy}$ by
\be
\theta_K=-\Im\left\{ \frac{\tilde{n}_R-\tilde{n}_L}{\tilde{n}_R\tilde{n}_L-1}\right\} = \frac{4\pi}{\omega}\Im \left\{ \frac{1}{\bar{n}(\bar{n}^2-1)}\sigma_{xy} \right\} 
\ee
where $\omega$ is the frequency of the incident light and $\bar{n}(\omega)$ is the average complex index of refraction  for the material \cite{Mineev2007}. A small but finite PKE is associated with the onset of superconductivity in the putative $p_x\pm ip_y$ spin-triplet superconductor \SRO~\cite{Xia2006b}, as well as in the low-temperature $B$ phase of the multiphase  heavy-fermion superconductor UPt$_3$, where the proposed $E_{2u}$ order parameter breaks TRS in the $B$ phase but not in the higher-temperature $A$ phase \cite{Schemm2014}. Theoretical  studies suggest that on a microscopic level, such an effect may arise extrinsically from impurity scattering \cite{Goryo2008,Lutchyn2009} or intrinsically from interband interactions \cite{Taylor2012}, but in either case is driven by the underlying symmetry of the superconducting state.

In this Letter we report the results of polar Kerr effect measurements performed at low temperatures on multiple high quality single crystals of \URS. We find a small ($\sim$160 nanorad at 300 mK), sample-independent, field-trainable Kerr effect that onsets near $T_c\sim1.5$ K. The general behavior of the measured signal is consistent with a TRSB order parameter intrinsic to the superconducting state with some component along $\hat{c}$. However, an additional feature is observed at $\sim$0.8-1 K that is unique among systems studied using this technique, suggestive of physics beyond the simplest $E_g$ picture of superconductivity in \URS. In contrast to the superconducting state, a finite signal that appears in the vicinity of the hidden order transition for samples trained in an external magnetic field may point to extrinsic effects. 

\bigskip

\begin{figure}
\includegraphics[width=\columnwidth]{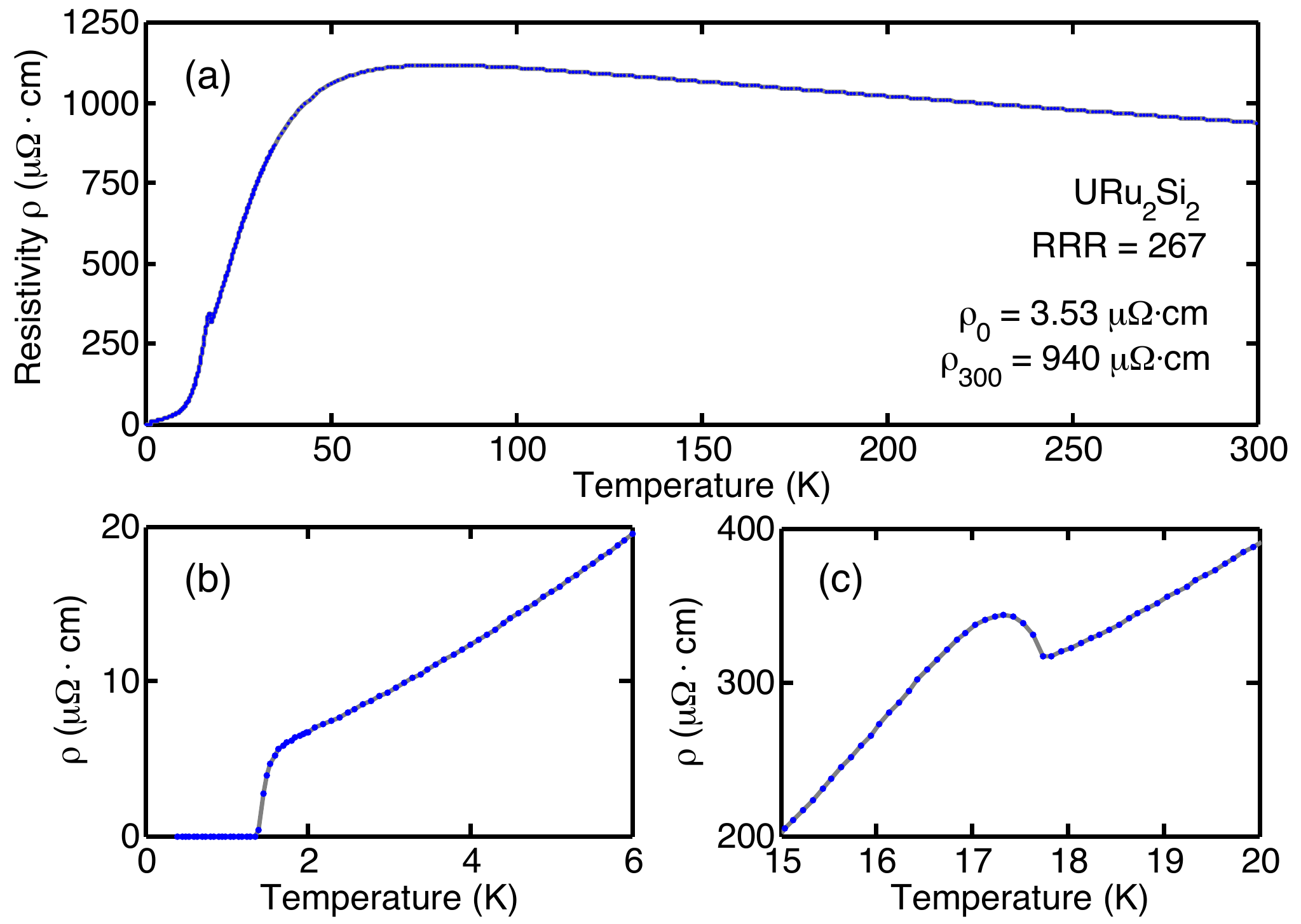}
\caption{Resistivity data for the RRR = 267 \URS~single crystal studied in this work. Analogous characterization of the two other samples yielded RRR values of 318 and 453. (a) The high temperature data shows the broad Kondo peak associated with formation of heavy charge carriers. The transitions at (b) $T_c \text{(midpoint)} = 1.49$ K and at (c) $T_{HO}  = 17.5$ K are clearly resolved.}
\label{RvsT}
\end{figure}

Single crystals of \URS~were grown by the Czochralski technique and electro-refined to improve purity. The crystals were oriented  by Laue diffraction in the back-scattering geometry and were cleaved or cut with a wire saw. Each resulting  $\sim$1$\times$0.3$\times$0.2 mm$^3$ sample of \URS~was characterized by electrical resistivity and can be parametrized by its residual resistivity ratio RRR $ = \rho(300 {\rm K})/\rho(0 {\rm K})$, where $\rho(0)$ was obtained  from a power law fit of the form $\rho(T) = \rho_0 + AT^n$ at low temperatures. For the crystal with RRR = 267, this analysis yielded values of $\rho_0  = 3.53$ $\mu\Omega\cdot$cm and $n = 1.5$ (Fig.~\ref{RvsT}); RRR values of 318 and 453 were found in a similar manner for the remaining two samples. The midpoints of the resistive transitions for the RRR = 267, 318, and 453 samples were found to be 1.49 K, 1.52 K, and 1.54 K, respectively. The crystals were then manually aligned and affixed with thermal grease to a copper stage bolted to the cold finger of a commercial $^3$He cryostat for Kerr effect measurements.

The Kerr angle $\theta_K$ was measured to high accuracy ($\pm20$ nanorad) using a zero-area loop Sagnac interferometer similar to that described in \cite{Xia2006a}, operating at a wavelength of 1550 nm with spot size $D\sim10.6$ $\mu$m. The incident optical power was kept to $P_{\text{inc}}\approx 20$ $\mu$W in order to minimize any effects of sample heating. External magnetic fields, when applied, were oriented along the $c$-axis during cooldown and switched off at base temperature (300 mK). All of the data presented in this paper were taken upon warmup from 300 mK in zero applied field; a Hall bar mounted in the sample space verified the residual field under these conditions to be less than  $\sim$3 mG. A series of measurements was also performed in a double $\mu$-metal shielded environment to achieve true ($<0.3$ mG) zero-field conditions.   

\begin{figure}
\includegraphics[width=\columnwidth]{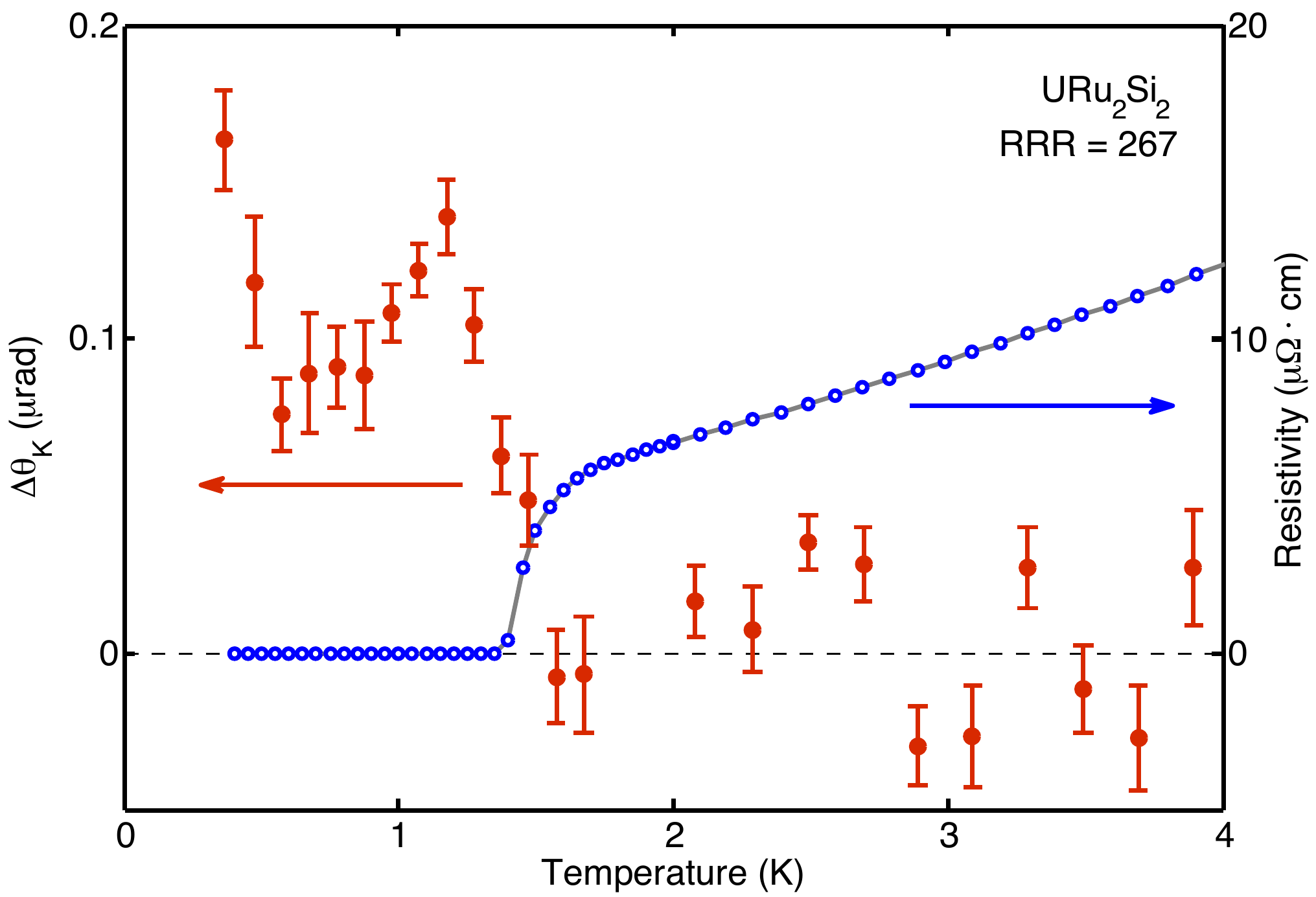}
\caption{Kerr effect data taken on a \URS~single crystal with RRR = 267 upon warmup in a true zero field ($\mu$metal  shielded) environment. Right axis:~resistivity data (open blue points) from Fig.~\ref{RvsT} showing the superconducting transition at $T_c\text{(midpoint)}  = 1.49$ K. Left axis:~change in Kerr angle (filled red circles) relative to the background at 5 K, showing an onset of Kerr signal at $T_c$ and a sharp anomaly at $\sim$1 K. Error bars in the Kerr effect data are statistical.}
\label{KerrSC}
\end{figure}

Our main experimental result is shown in Fig.~\ref{KerrSC} for a crystal with RRR = 267. The Kerr effect as a function of temperature is plotted on the left axis relative to a small, temperature-independent (below $\sim$5 K) background which will be further discussed below. Each point represents an average of approximately 600 data points, and the corresponding error bars are statistical. For comparison, resistivity data from the same crystal is replotted from Fig.~\ref{RvsT} on the right axis, showing the transition to the superconducting state with $T_c\text{(midpoint)} = 1.49$ K.
 
There are two significant features in the PKE data of Fig.~\ref{KerrSC}. At base temperature, an additional Kerr rotation of approximately 160 nanorad is clearly resolved relative to the 5 K background, which decays to the background value at $T_c$ to within experimental resolution. This overall behavior, including the sharp onset at $T_c$, is consistent with the development of a TRSB order parameter through a second-order phase transition and is reminiscent of the PKE associated with the chiral superconducting states in \SRO~and UPt$_3$. The size of the PKE is similar to that found in UPt$_3$~\cite{Schemm2014}, and is independent of the RRR of the samples. This sample independence, together with the high RRR values of the crystals measured in this study, suggests that the observed PKE most likely originates from multi-band superconductivity itself \cite{Taylor2012} rather than from impurity scattering \cite{Goryo2008,Lutchyn2009}. In addition, we note a pronounced anomaly in the PKE data at $T\sim1$ K. This unusual feature has been observed in every PKE measurement of every \URS~sample taken to date and will be returned to in the discussion.

A key signature of a second-order phase transition is the sensitivity of the order parameter to cooling through the transition temperature in the presence of a symmetry-breaking field. In the absence of such a training field, one generically expects the ordered state to form domains, with the direction (sign) of the ordered moment varying between them. Thus, if the beam diameter of the Sagnac probe is larger than the average domain size, the net observed signal will vary in sign and in magnitude between cooldowns. Conversely, the application of a small external field as $T_c$ is crossed will train the sign of the moment across the sample, leading to a fully saturated signal aligned with the direction of the training field. 

\begin{figure}
\includegraphics[width=\columnwidth]{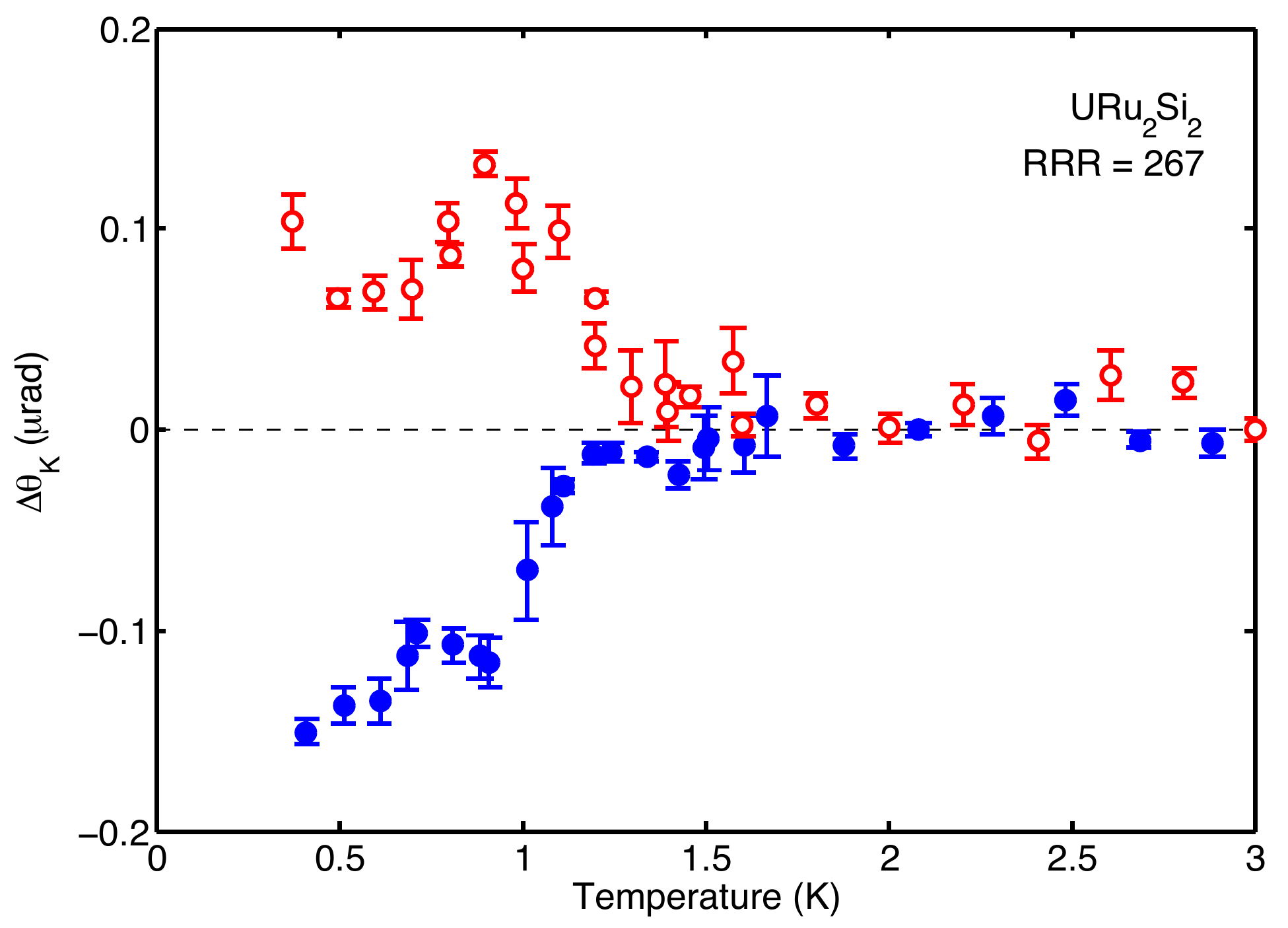}
\caption{Field training of PKE in the superconducting phase of \URS. Open red circles: zero-field warmup data following +100 G cooldown in the same crystal. Filled blue circles: zero-field warmup data following -100 G cooldown, showing complete reversal of the Kerr signal.}
\label{KerrField}
\end{figure}

In the case of a chiral superconductor, an order parameter of the form (\ref{gap}) is expected to couple to external magnetic fields applied along the $c$-axis. Fig.~\ref{KerrField} shows the results of PKE measurements taken in zero field after cooling the RRR = 267 sample to base temperature in the presence of a small ($\pm$100 G) $c$-axis oriented field. The field trained signals are equal and opposite in sign to within experimental resolution, as one would expect based on the picture outlined above. In addition, the magnitude of the field cooled signals matches that of the largest zero field cooled signals, which has two additional implications. First, while domain size may vary, there are domains that are of the size of the beam, or even larger. Second, when measuring any type-II superconductor one must be aware of the potential contribution of trapped vortices to PKE. The fact that both the magnitude and the form of the field cooled Kerr signal match the zero field ($\mu$-metal shielded) maximum, and that this observation is independent of the RRR of the sample, indicates that trapped flux is not the source of the observed PKE, suggesting that this signal is intrinsic to the superconducting state. 

\begin{figure}
\includegraphics[width=\columnwidth]{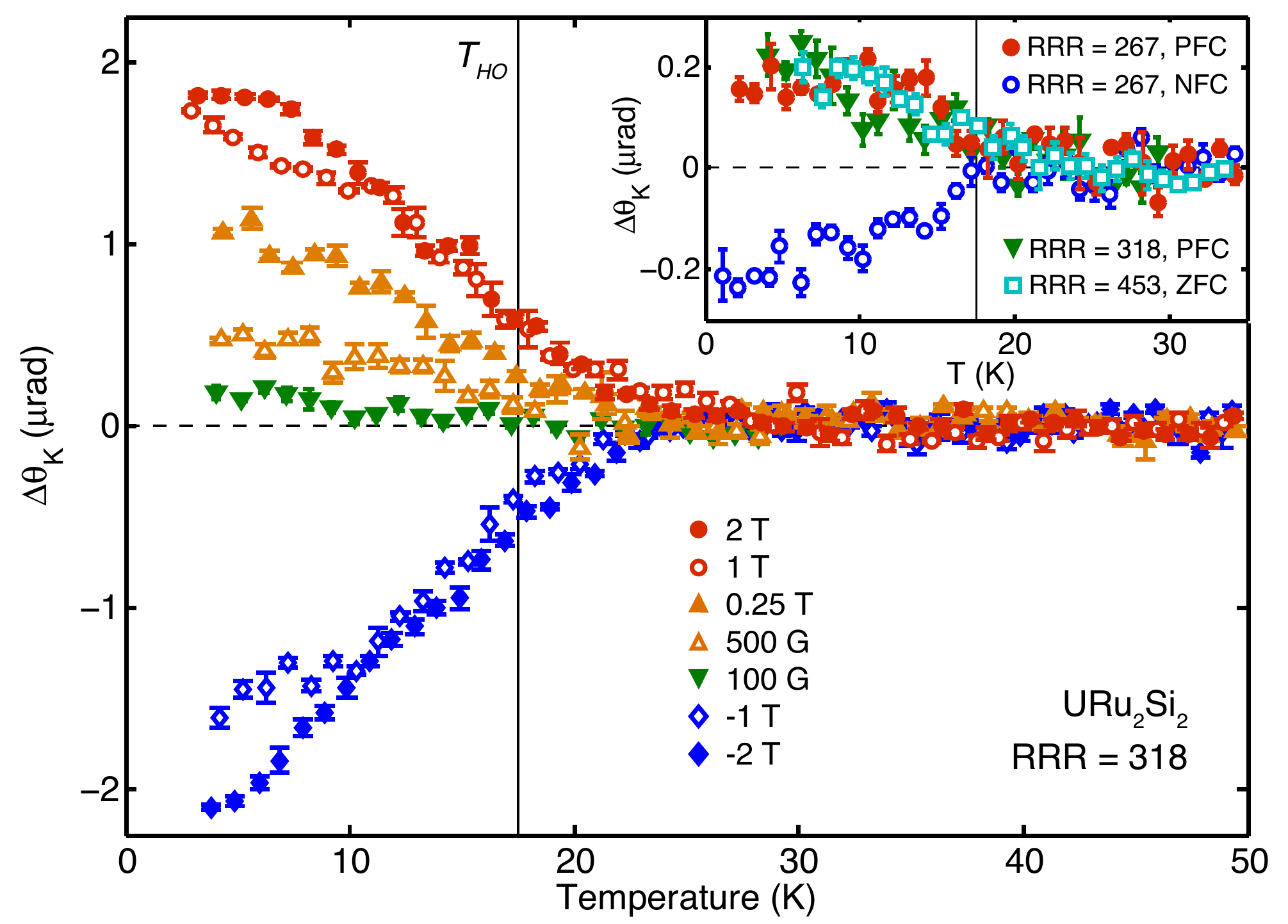}
\caption{Kerr effect in the hidden order phase. Main panel: Kerr effect in the HO r\'egime following high field training for the RRR = 318 crystal. The remanent Kerr rotation depends on the magnitude of the training field, up to about 1 T, in contrast to the field training response expected for TRSB developing through a second-order phase transition. The collection of $\theta_K(T)$ curves does not converge until well above $T_{HO}$. Inset: A small, field-trainable Kerr rotation onsets in the vicinity of $T_{HO} = 17.5$ K in all samples. This signal saturates by $T=5$ K, permitting a mostly independent examination of PKE in the superconducting state relative to this background. PFC = positive (+100 G) field-cooled, NFC = negative (-100 G) field-cooled, ZFC = zero-field-cooled.}
\label{HOField}
\end{figure}

A second series of questions concerns the unique hidden order state from which superconductivity emerges in \URS. Of particular interest to this study is whether a finite (intrinsic or extrinsic) PKE develops in the hidden order phase, and if so whether it influences the TRSB signal in the superconducting phase. A plot of Kerr rotation versus temperature up to 35 K, under the field training conditions relevant to this study, is shown in the inset of Fig.~\ref{HOField}. It is apparent that \URS~already exhibits PKE well above $T_c$. This high-temperature Kerr signal onsets in the vicinity of the hidden order transition and saturates with a magnitude not larger than $|\Delta \theta_K^{\rm sat}|\sim 200$ nanorad as $T\rightarrow 0$ K for all samples studied. It is important to note that this `normal state' Kerr response is insufficient to account for the PKE in the superconducting state via vortex core contributions: given that $H_{c2} \sim 1.5$ T and the maximum field used is 100 G, the largest possible contribution to $\theta_K$ from vortex core formation is $\Delta\theta_K^{\rm vor} = \Delta \theta_K^{\rm sat}\cdot \left(\frac{100}{15000} \right) \sim 1.3$ nrad. Furthermore, the relationship between this signal and the hidden order state is unclear, since cooling the sample in increasingly larger magnetic fields not only increases the size of the residual signal (which is always measured during zero field warmup), but also clearly indicates that the onset temperature, $T_\mathrm{onset} \sim$ 25 K, well exceeds $T_{HO} = 17.5$ K. 

To determine whether the unusual high temperature data of Fig.~\ref{HOField} have any relationship to either the sign or the magnitude of the PKE data in the superconducting phase, a series of experiments was performed in which the sample was cooled from $\sim$35 K to 300 mK in a small ($H_\mathrm{app} = \pm 100$ G) field. At 300 mK the field was switched off and $\theta_K$ was measured during warmup to $\sim$5 K $\ll 17.5$ K. At $\sim$5 K the magnetic field was reapplied but with opposite sign, the sample was re-cooled to base temperature, and $\theta_K$ was again measured in zero field during warmup.

\begin{figure}
\includegraphics[width=\columnwidth]{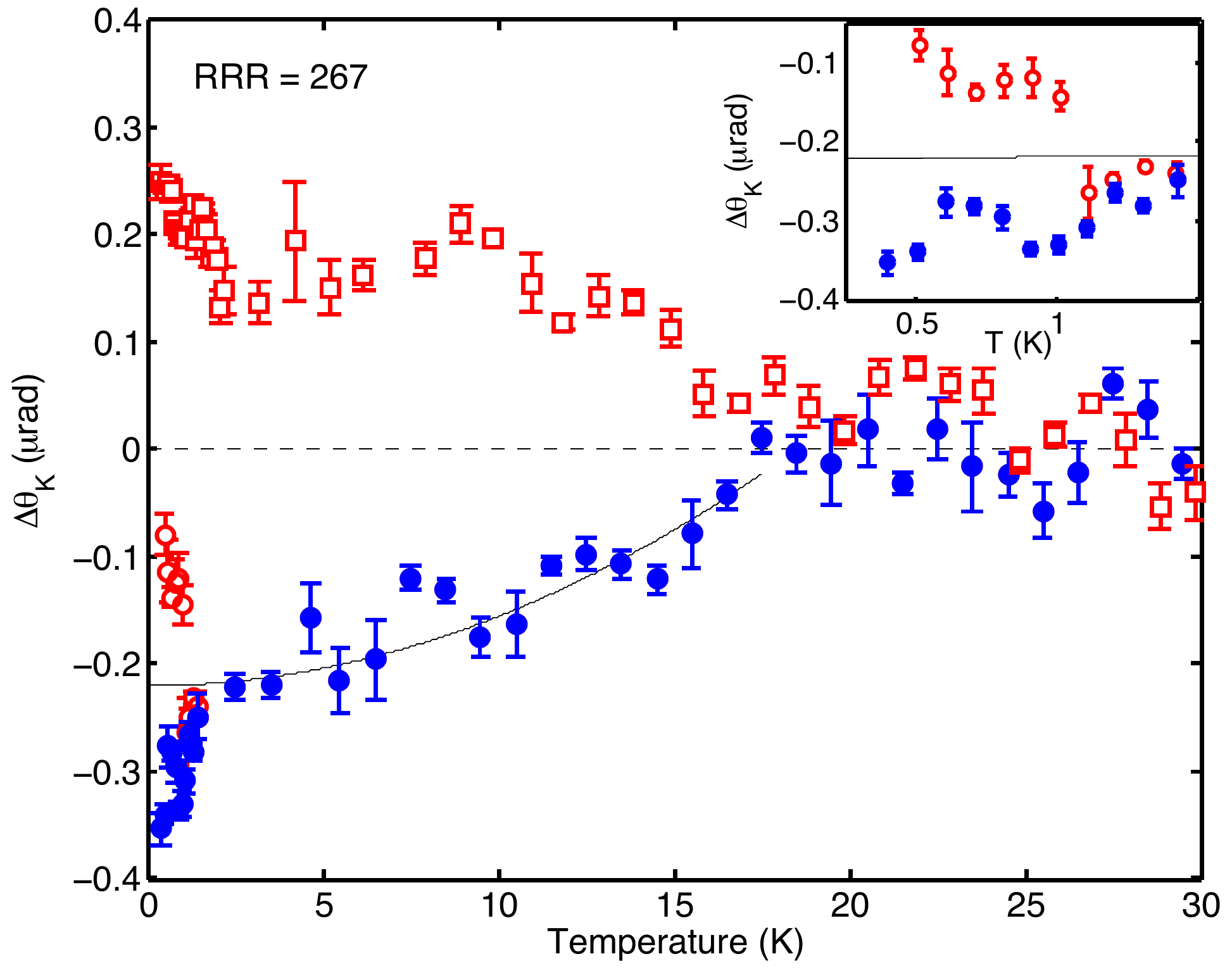}
\caption{Independent field training of the hidden order and superconducting PKE signals in \URS~for the sample with RRR = 267. Red and blue (open and filled symbols) indicate whether a positive or negative training field was applied during cooldown through each temperature range. Solid line is a guide to the eye of the form $\bigl(1-(T/T_\mathrm{onset})^2 \bigr)$. Inset: magnification of the lower left portion of the main panel data, showing more clearly the reversal of Kerr signal, including the $\sim$1 K feature, in the superconducting state, with no change in background.} 
\label{HOSC}
\end{figure}

Fig.~\ref{HOSC} shows the result of one such series of measurements from the crystal with RRR = 267. Similar behavior was also observed, albeit with less sensitivity, in the sample with RRR = 318. All of the data in the figure are plotted relative to backgrounds taken between 30 and 35 K toward the end of their respective full warmups. It is apparent from the figure that both the superconducting and $\sim$25 K signals are trained in the direction of the applied field, when the field is applied throughout cooldown to 300 mK. However, if the training field direction is reversed at 5 K -- above $T_c$ but below $T_\mathrm{onset}$ -- the superconducting contribution to the signal is reversed in sign with no change in magnitude, while the high temperature signal -- upon which the superconducting features are superimposed -- remains unchanged. In other words, the high temperature signal is influenced only by the training field applied through $\sim$25 K, while the additional low temperature signal is influenced only by the training field applied through $T_c$. This independence suggests that the two signals are unrelated; otherwise, the high temperature signal would be strong enough to overcome a training field of the opposite sign and determine the TRSB sign of the superconducting phase.


\bigskip

The apparent independence of the high- and low-temperature PKE signals, along with the atypical field training response of the high-temperature signal, suggest that the two effects may have different origins. \URS~is often found to be inhomogeneous, presumably due to strains resulting from local crystalline defects or applied external pressure, resulting in a minority phase composed of large magnetic moments that coexists within a majority (bulk) nonmagnetic hidden order phase \cite{Matsuda2001,Takagi2007}. It is possible that the PKE observed beginning at $\sim$25 K is due to these contaminants and is not intrinsic to the HO phase. We note, however, that this interpretation only implies that the HO phase does not exhibit TRSB resulting in a net moment along the $c$-axis. TRSB ordering whose moment along the $c$-axis averages to zero within the (relatively large) area of the Sagnac probe beam may still be present, although an in-plane moment, which should not give rise to PKE in our experimental geometry, was determined via neutron scattering measurements to be not more than $\sim$1$\times$$10^{-3}$ $\mu_B$/U \cite{Das2013, Ross2014}.

The additional PKE observed below $T_c$ is consistent across samples and is field trainable independently of the high temperature signal. We therefore believe that the observed TRSB signal reflects an intrinsic property of superconductivity in \URS. Such an intrinsic signal is consistent with the $k_z(k_x\pm ik_y)$ order parameter symmetry proposed in \cite{Kasahara2007} and is supported by recent $\mu$SR data indicating an internal magnetic moment in the superconducting phase \cite{Kawasaki2014}. Assuming a single superconducting phase in the entire $H$-$T$ phase diagram, our observations stand in contrast to the recent report that the superconducting state possesses an {\em in-plane} TRSB moment, with little to no component along $\hat{c}$ \cite{Li2013}. The conclusions of \cite{Li2013} are based on magnetic torque measurements in very high (up to 18 T) fields, however, and therefore may point to a different superconducting phase at finite field.

The PKE anomaly at $\sim$1 K presents a particular challenge to understanding superconductivity in \URS. The data in this Letter are not the first indication of unusual behavior within the superconducting state. A vortex lattice melting transition produces a peak in $d\rho / dT$ below $T_c$ as well as a peak-dip feature in the temperature-normalized thermal conductivity $\kappa /T$ \cite{Okazaki2008, OkazakiJPhys2009} -- although such features require that there be a magnetic field present to nucleate vortices. At the same time, measurements of $H_{c1}$ ($\mathcal{O}(10)$ Gauss) for $H \, ||\, \hat{c}$ show a kink feature at $\sim$1--1.2 K that (1) appears to be intrinsic, and (2) cannot be fit with a simple two-band model for the gap \cite{Okazaki2011}.

This collection of anomalous observations leaves open the possibility of a novel pairing mechanism in superconducting \URS. In one such picture, a mixed singlet-triplet $d$-density wave (st-DDW) with `skyrmionic' spin texture may lead to chiral $d$-wave superconductivity \cite{Hsu2013} through either of two ($\ell = 2$ or $\ell = 4$) angular momentum channels; if both channels are represented, a double transition would take place \cite{Hsu2013}. A second theory involving spin-textured states with $\ell = 2$ admits both $m_z = \pm 1$ (`Weyl fermions', $\Delta \sim k_z(k_x\pm ik_y)$) and $m_z = \pm2$ (`double-Weyl fermions', $\Delta \sim k_z(k_x\pm ik_y)^2$) states \cite{Goswami2013}. These are topologically distinct, would both generate PKE, and could also give rise to a double transition. A third proposal posits that, as in UPt$_3$, the real and imaginary components of the order parameter may not have the exact same onset temperature \cite{Okazaki2011}, although in this case no PKE should be generated until the lower of the two superconducting transition temperatures \cite{Schemm2014}. Further measurements will be required to determine which, if any, of these pictures accurately account for the superconducting behavior in \URS.

\acknowledgments
Stimulating discussions with Sudip Chakravarty, Pavan Hosur, Steve Kivelson, Joseph Orenstein, Srinivas Raghu, and Chandra Varma are greatly appreciated. Sample preparation and characterization at LANL were supported by the U.~S.~Department of Energy (DOE) Office of Basic Energy Science, Division of Materials Science and Engineering; Kerr effect measurements at Stanford were supported under DOE contract No.~DE-AC02-76SF00515. Construction of the Sagnac apparatus was partially supported by the NSF through Stanford's Center for Probing the Nanoscale.

\end{document}